\documentclass{sf2a-conf2019}
\usepackage{graphicx}
\usepackage{hyperref}
\usepackage[]{natbib}  
\usepackage{epstopdf}
\usepackage[varg]{txfonts}
\usepackage{color}
\usepackage{rotating}

\def\BibTeX{{\rm B\kern-.05em{\sc i\kern-.025em b}\kern-.08em
    T\kern-.1667em\lower.7ex\hbox{E}\kern-.125emX}}
\bibpunct{(}{)}{;}{a}{}{,}  

\newcommand{\ra}{\rightarrow}

\begin{document}

\TitreGlobal{SF2A 2019}

\title{Changing-look Seyfert galaxies with optical linear polarization measurements}

\runningtitle{The optical polarization of changing-look Seyferts}

\author{F.~Marin}\address{Universit\'e de Strasbourg, CNRS, Observatoire Astronomique de Strasbourg, UMR 7550, F-67000 Strasbourg, France}

\author{D.~Hutsem\'ekers}\address{Institut d'Astrophysique et de G\'eophysique, Universit\'e de Li\`{e}ge, All\'ee du 6 Ao\^{u}t 19c, 4000 Li\`{e}ge, Belgium}

\author{B.~Ag\'is Gonz\'alez$^{2}$}

\setcounter{page}{237}

\maketitle

\begin{abstract}
In this lecture note, we make the case for new (spectro)polarimetric measurements of ``changing-look'' AGNs (CLAGNs),
a subclass of the AGN family tree that shows long-term (months to years) large flux variability associated with 
the appearance or disappearance of optical broad emission lines. We discuss how polarization measurements could 
help to distinguish which of the several scenarios proposed to explain such variations is/are the most likely.
We collected past polarization measurements of nearby, Seyfert-like CLAGNs and take stock that almost 
all polarimetric information we have on those fascinating objects dates from the 80's and 90's. We thus explain
how polarization could help us understand the physical processes happening in the first parsecs of CLAGNs and
why new polarization monitoring campaigns are strongly needed.
\end{abstract}

\begin{keywords}
Galaxies: active, galaxies: quasars, galaxies: Seyfert, polarization
\end{keywords}


\section{Introduction}
\label{introduction}
Among active galactic nuclei (AGNs), a new class of object is nowadays recognized. Those specific AGNs have time-dependent 
spectroscopic signatures that makes them appear as type-1 AGNs for a certain period and then as type-2 AGNs after a while
(see, e.g., \citealt{1971Khachikian,1986Cohen,1989Goodrich}). Type-1 AGN are characterized by large optical fluxes 
associated with broad ($>$ 1000~km.s$^{-1}$) and narrow ($\le$ 1000~km.s$^{-1}$) emission lines, while type-2s only show
narrow emission lines and lower optical fluxes. The large Doppler widths result from photo-ionization of an equatorial 
reservoir of gas composed of many cloudlets that have large Keplerian velocities and densities \citep{2009Gaskell}. 
Depending on the inclination of the system with respect to the observer, this broad emission line region (BELR) may be 
hidden by an optically thick, equatorial, circumnuclear layer of dust. This orientation dependence has been used to explain
the observational differences between the two AGN types for decades now and is still a very robust interpretation 
\citep{1993Antonucci}. However, there are rogue AGNs that have shown type transitions on timescales of 
months to years. Examples of such objects are Mrk~1018, which varied between type-2 and type-1 between 1979 and 1984 
\citep{1986Cohen}, NGC~4151 that changed from type-1 to type-2 between 1974 and 1984 \citep{1984Penston}, or 3C~390.3 
that followed the same type transition between 1975 and 1984 \citep{1984Penston}. From dynamical timescale arguments, 
it is physically impossible that a parsec-sized object has changed its whole inclination in a human time frame. Then, 
how can we explain those ``rapid'' changes of type ? There are several theories involving the appearance or disappearance  
of optically thick material in front of the observer's line-of-sight \citep{1989Goodrich,2012Elitzur}, tidal disruptions events 
(TDEs, \citealt{1988Rees,2016Lawrence}) or rapid mass accretion rate drop resulting in the disappearance of the BELR 
\citep{2018Noda}. In this lecture note, we will expound how (spectro)polarimetric measurements of those ``changing-look AGN'' 
(CLAGNs) could help understand the physical processes happening around active supermassive black holes. Here, we will 
focus on nearby, low-luminosity CLAGNs, and refer to \citet{2019Hutsemekers} for high luminosity objects (quasars).

\section{Optical polarization of Seyfert-like CLAGNs}
\label{CLAGNs}
There are at least three scenarios\footnote{We also take note of the controversial explanation of large amplitude microlensing 
by stars in foreground galaxies to explain CLAGNs \citep{2016Lawrence}} to explain the dramatic flux variation and 
spectral change of Seyfert-like CLAGNs. The first one invokes the appearance or disappearance of obscuring material 
in front of the observer's line of sight. In this case, a cloud from either the outer BELR or the circumnuclear torus 
passes in front of the line-of-sight and (partially) obscures the central source, resulting in an opacity-dependent 
dimming and the apparent disappearance of the broad emission line signatures \citep{1989Goodrich,1992Tran}. The second 
scenario explains CLAGNs using unusually luminous TDEs \citep{2016Lawrence}. When a star orbits close enough
to the central supermassive black hole of AGNs, it is torn apart by tidal forces and a fraction of the mass is
accreted, resulting in a sudden brightening of the black hole. The change in luminosity can easily last 
for several hundreds of days \citep{1988Rees}. Finally, a third scenario postulates that the CLAGN phenomenon 
is due to modifications in the source of ionizing radiation, likely a variation in the rate of accretion onto the 
central supermassive black hole \citep{1984Penston,2014Elitzur,2018Noda}.

Spectroscopic and photometric observations can be explained by one or several of those scenarios, depending on 
the target. However, their polarization signatures are unique \citep{2016Marin,2017Hutsemekers,2017Marin,2019Hutsemekers}.

\begin{itemize}
\item If the central source is intrinsically dimming, at the onset of the flux variation the polarization degree experiences 
sharp decreases and increases associated with rotations of the polarization angle. Those time-dependent variations are due to 
lower amounts of direct, unpolarized flux from the central engine and constant amounts of reprocessed (delayed) radiation from 
the equatorial region. The duration of the high polarization degree peak depends on the distance of the scatterer from the source 
and can be used to achieve polarized reverberation mapping of the inner CLAGN regions. The polarization degree and polarization
position angle then return to a stability period after several years/decades (see Marin \& Hutsem\'ekers, A\&A, submitted). 
If the BELR disappears, electron scattering inside the BELR becomes inefficient, the polarization degree decreases and the 
polarization position angle rotates by 90$^\circ$. Polarized light echoes are much less bright due to the absence of an 
electron-filled, nearby scattering region. The duration of the echo is also longer due to the fact that radiation has to scatter
onto the parsec-scale torus/winds rather than onto the sub-parsec scale BELR. At the end of the echo, the polarization position 
angle rotates again by 90$^\circ$, returning to the initial value at the same time that the polarization degree returns to a 
stability period. This could, in turn, provide us with an estimate of the inner radius of the torus/wind if the polarized light echo
is detectable.

\item In the case of cloud obscuration, radiation mainly escapes the central (obscured) region by scattering inside the polar 
outflows, similarly to what has been postulated for the Unified Scheme of AGNs \citep{1993Antonucci}. This results in much higher
polarization degrees (10 -- 20\%, see e.g., NGC~1068 \citealt{1985Antonucci}) and a rotation of the polarization position angle
due to the fact that equatorial scattering is no longer visible. The flux and polarization variations are also time-dependent but
are likely to be shorter depending on the size and radial distance of the cloud to the central engine \citep{2018Gaskell}.
\end{itemize}

\begin{sidewaystable*}
\caption{Changing-look Seyferts with optical linear polarization measurements}
\centering
\label{Tab:CLAGN}
\begin{tabular}{lcccc}
\hline\hline
Object & Spectral type (year) & References & Polarization degree (year) & References \\
& & & (\% ) & \\
\hline
Mrk~6     & 2   {\small (1968)} $\ra$ 1.5 {\small (1969-2013)} & 1,2,3  & 0.54$\pm$0.15 {\small (1976)} $\ra$ 0.90$\pm$0.03 {\small (1997)} $\ra$ 0.74$\pm$0.17 {\small (2013)} & 4,5,3 \\
Mrk~372   & 1.5 {\small (1986)} $\ra$ 1.9 {\small (1990)} & 6 & 1.49$\pm$0.46 {\small (1976)} & 4  \\
Mrk~590   & 1.5 {\small (1973)} $\ra$ 1 {\small (1989-1996)} $\ra$ 1.9  {\small (2006-2014)} $\ra$ 1  {\small (2017)}  &  7,8  & 0.32$\pm$0.30 {\small (1976)} & 4   \\
Mrk~1018  & 1.9 {\small (1979)} $\ra$ 1 {\small (1984-2009)} $\ra$ 1.9  {\small (2015)} &  9,10   & 0.28$\pm$0.05 {\small (1986)} & 11  \\
NGC~1566  & 1   {\small (1962)} $\ra$ 1.9 {\small (1969)} $\ra$ 1  {\small (1980)} $\ra$ 1.9  {\small (1985)} $\ra$ 1  {\small (2018)} & 12,13,14 & 0.60$\pm$0.24 {\small (1980)}  $\ra$ 1.33$\pm$0.18 {\small (1997)} & 4,15  \\
NGC~2617  & 1.8 {\small (1994-2003)} $\ra$ 1 {\small (2013-2016)}  &  16,17,18 & 0.43$\pm$0.15 {\small (1998)} & 19  \\
NGC~2622  & 1.8 {\small (1981)} $\ra$ 1 {\small (1985-1987)} &  11  & 2.35$\pm$0.03 {\small (1986)} & 11  \\
NGC~3516  & 1   {\small (1996-1998)} $\ra$ 1 {\small (2007)} $\ra$ 2 {\small (2014-2017)} & 20 & 0.15$\pm$0.04 {\small (1997)} & 5  \\
NGC~4151  & 1   {\small (1974)} $\ra$ 1.9 {\small (1984-1989)} $\ra$ 1.5 {\small (1990-1998)} $\ra$ 1.8 {\small (2001)} &  21,22,23  & 0.26$\pm$0.08 {\small (1976)} $\ra$ 1.18$\pm$0.05 {\small (1992)} $\ra$ 0.32$\pm$0.30 {\small (2014)} & 4,24,25  \\
NGC~7582  & 2   {\small (1980-1998)} $\ra$ 1 {\small (1998)}  &  26,27  & 1.03$\pm$0.12 {\small (1981)} & 4  \\
NGC~7603  & 1   {\small (1974)} $\ra$ 1.8 {\small (1975)} $\ra$ 1  {\small (1976-1998)} &  28,29 & 0.32$\pm$0.29 {\small (1976)}  $\ra$ 0.42$\pm$0.03 {\small (1987)} $\ra$ 0.25$\pm$0.04 {\small (1997)} & 4,11,5  \\
Fairall~9 & 1  {\small (1977-1981)} $\ra$ 1.8 {\small (1984)} $\ra$ 1 {\small (1987)} &  30,31  & 0.40$\pm$0.11 {\small (1981)} $\ra$ 0.37$\pm$0.13 {\small (1997)} & 4,5  \\
3C~390.3  & 1   {\small (1975)} $\ra$ 1.9 {\small (1980-1984)} $\ra$ 1 {\small (1985-1988)} $\ra$ 1 {\small (2005-2014)}  &  21,32,33 & 0.84$\pm$0.30 {\small (1976)}  $\ra$ 1.30$\pm$0.10 {\small (1986)} $\ra$ 1.13$\pm$0.18 {\small (2014)} & 4,34,35  \\
\hline
\end{tabular}
\begin{flushleft}
{(1)~~\citet{1971Khachikian}; (2)~\citet{2011Khachikian}; (3)~\citet{2014Afanasiev}; (4)~\citet{1983Martin}; (5)~\citet{2002Smith}; (6)~\citet{1991Gregory}; (7)~\citet{2014Denney}; (8)~\citet{2019Raimundo}; (9)~\citet{1986Cohen}; (10)~\citet{2016McElroy}; (11)~\citet{1989Goodrich};  (12)~\citet{1970Pastoriza}; (13)~\citet{1986Alloin}; (14)~\citet{2019Oknyansky}; (15)~\citet{1999Felton}; (16)~\citet{1996Moran}; (17)~\citet{2014Shappee}; (18)~\citet{2017Oknyansky}; (19)~\citet{2011Wills}; (20)~\citet{2019Shapovalova}; (21)~\citet{1984Penston}; (22)~\citet{1997Malkov}; (23)~\citet{2008Shapovalova}; (24)~\citet{1998Martel}; (25)~\citet{2019Afanasiev}; (26)~\citet{1980Ward}; (27)~\citet{1999Aretxaga}; (28)~\citet{1976Tohline}; (29)~\citet{2000Kollatschny}; (30)~\citet{1985Kollatschny}; (31)~\citet{1992Lub}; (32)~\citet{1991Veilleux}; (33)~\citet{2017Sergeev}; (34)~\citet{1991Impey}; (35)~\citet{2015Afanasiev}
}\\\end{flushleft}
\begin{flushleft}
{$ $ \\
{{\it Mrk~6} : Variations in type 1.5 state on short timescales. Equatorial scattering dominated \citep{2004Smith}. Polarization reverberation \citep{2014Afanasiev}. No  strong variation of PPA.}\\
{{\it Mrk~590} : Variations are likely intrinsic. Candidate for polarization echoes.}\\
{{\it Mrk~1018} : Variations are likely intrinsic. Candidate for polarization echoes. }\\
{{\it NGC~1566} : Recurrent variations with outbursts from type 1.9/1.8 to type 1.5/1.2 during decades. No strong variation of PPA.}\\
{{\it NGC~2617} : Variations are likely intrinsic. Candidate for polarization echoes. }\\
{{\it NGC~2622} : Variations are likely due to obscuration. Polar scattering dominated \citep{2004Smith}.}\\
{{\it NGC~3516} : Complex variability from 1999 to 2008. Variations are likely due to obscuration.}\\
{{\it NGC~4151} : Complex variability on multiple timescales.   Equatorial scattering dominated \citep{2004Smith}. Polarization reverberation \citep{2012Gaskell}. No strong variation of PPA.}\\
{{\it NGC~7582} : The transition to type 1 was fast, and short.}\\
{{\it NGC~7603} : Variations are likely due to obscuration.}\\
{{\it 3C 390.3} : Complex variability. Radio galaxy. Polarization reverberation \citep{2015Afanasiev}.}\\
}\end{flushleft}
\end{sidewaystable*}

All expected differences are detailed in \citet{2017Marin}. In any case, it is vital to obtain polarization measurements 
of CLAGNs, before and after the change of look. We thus compiled the historical spectral type changes and polarization
measurements of known changing-look Seyferts (at our best knowledge) in Tab.~\ref{Tab:CLAGN}. The spectral types of 
changing-look Seyferts and the epoch at which they were measured are given in Col.~2. A range of dates indicates that
the spectral types measured at these two dates are identical, with no change recorded in between. We emphasize that 
this does not imply absence of spectral type variations during this period due to a lack of data. For some objects, 
exhaustive monitorings were carried out. In such cases, only some representative types/epochs are reported in 
Tab.~\ref{Tab:CLAGN}. The polarization degrees given in Col.~4 refer to the optical continuum polarization measured 
in various broad-band filters. For a few objects the polarization was monitored during several years. In such cases,
we give three representative values at most in Tab.~\ref{Tab:CLAGN}. In total, there are only 23 representative 
polarization measurements of Seyfert-like CLAGNs. Among the 23, only 3 observations have been carried out after 2000,
which means that our knowledge of the polarization of CLAGNs is based on data that are at least 20 years old. There 
are only 6 objects (Mrk~6, NGC~1566, NGC~4151, NGC~7603, Fairall~9 and 3C~390.3) that have repeated polarimetric
measurements but none of them happen to coincide with the change of look. At best, we can estimate the past 
polarization level of CLAGNs before their transition but there is very little we can do about determining the correct
physical explanation of the spectral/flux change without new and regular polarization measurements of those objects.

\section{Discussion and conclusions}
\label{conclusion}
We have seen that the pool of archival polarimetric measurements of state transitions in CLAGNs is very limited, 
almost non-existent. This is rather detrimental since polarimetric observational data along with numerical models 
are a unique tool to determine what are the physical causes of the changes of look, unveiling new frontiers in 
the AGN physics. New and repeated polarimetric measurements are thus needed as part of a monitoring campaign. 
In addition to the Seyfert-like CLAGNs discussed here, there are at least 23 candidates for a follow up program 
and a handful more of radio-loud AGNs \citep{2019Hutsemekers}. Ideally, broad-band polarization measurements 
should be obtained twice a year during typically one or two decades. For the brightest objects (Seyferts) this 
could be achieved with robotic 1m class telescopes. On the other hand, a follow-up of the polarization of CL quasars
as those studied in \citet{2019Hutsemekers} would require 2-4m class telescopes, in particular when the objects 
are in their faint type-2 phase.

\begin{acknowledgements}
FM would like to thank the Centre national d'\'etudes spatiales (CNES) who funded his post-doctoral grant 
``Probing the geometry and physics of active galactic nuclei with ultraviolet and X-ray polarized radiative
transfer''. DH is senior research associate F.R.S.-FNRS. BAG acknowledges support provided by the Fonds de la 
Recherche Scientifique - FNRS, Belgium, under grant No. 4.4501.19.
\end{acknowledgements}

\bibliographystyle{aa}  
\bibliography{marin_S19}

\begin{thebibliography}{51}
\expandafter\ifx\csname natexlab\endcsname\relax\def\natexlab#1{#1}\fi

\bibitem[{{Afanasiev} {et~al.}(2019){Afanasiev}, {Popovi{\'c}}, \&
  {Shapovalova}}]{2019Afanasiev}
{Afanasiev}, V.~L., {Popovi{\'c}}, L.~{\v{C}}., \& {Shapovalova}, A.~I. 2019,
  \mnras, 482, 4985

\bibitem[{{Afanasiev} {et~al.}(2014){Afanasiev}, {Popovi{\'c}}, {Shapovalova},
  {Borisov}, \& {Ili{\'c}}}]{2014Afanasiev}
{Afanasiev}, V.~L., {Popovi{\'c}}, L.~{\v C}., {Shapovalova}, A.~I., {Borisov},
  N.~V., \& {Ili{\'c}}, D. 2014, \mnras, 440, 519

\bibitem[{{Afanasiev} {et~al.}(2015){Afanasiev}, {Shapovalova}, {Popovi{\'c}},
  \& {Borisov}}]{2015Afanasiev}
{Afanasiev}, V.~L., {Shapovalova}, A.~I., {Popovi{\'c}}, L.~{\v{C}}., \&
  {Borisov}, N.~V. 2015, \mnras, 448, 2879

\bibitem[{{Alloin} {et~al.}(1986){Alloin}, {Pelat}, {Phillips}, {Fosbury}, \&
  {Freeman}}]{1986Alloin}
{Alloin}, D., {Pelat}, D., {Phillips}, M.~M., {Fosbury}, R.~A.~E., \&
  {Freeman}, K. 1986, \apj, 308, 23

\bibitem[{{Antonucci}(1993)}]{1993Antonucci}
{Antonucci}, R. 1993, \araa, 31, 473

\bibitem[{{Antonucci} \& {Miller}(1985)}]{1985Antonucci}
{Antonucci}, R.~R.~J. \& {Miller}, J.~S. 1985, \apj, 297, 621

\bibitem[{{Aretxaga} {et~al.}(1999){Aretxaga}, {Joguet}, {Kunth}, {Melnick}, \&
  {Terlevich}}]{1999Aretxaga}
{Aretxaga}, I., {Joguet}, B., {Kunth}, D., {Melnick}, J., \& {Terlevich}, R.~J.
  1999, \apjl, 519, L123

\bibitem[{{Cohen} {et~al.}(1986){Cohen}, {Rudy}, {Puetter}, {Ake}, \&
  {Foltz}}]{1986Cohen}
{Cohen}, R.~D., {Rudy}, R.~J., {Puetter}, R.~C., {Ake}, T.~B., \& {Foltz},
  C.~B. 1986, \apj, 311, 135

\bibitem[{{Denney} {et~al.}(2014){Denney}, {De Rosa}, {Croxall}, {Gupta},
  {Bentz}, {Fausnaugh}, {Grier}, {Martini}, {Mathur}, {Peterson}, {Pogge}, \&
  {Shappee}}]{2014Denney}
{Denney}, K.~D., {De Rosa}, G., {Croxall}, K., {et~al.} 2014, \apj, 796, 134

\bibitem[{{Elitzur}(2012)}]{2012Elitzur}
{Elitzur}, M. 2012, \apjl, 747, L33

\bibitem[{{Elitzur} {et~al.}(2014){Elitzur}, {Ho}, \& {Trump}}]{2014Elitzur}
{Elitzur}, M., {Ho}, L.~C., \& {Trump}, J.~R. 2014, \mnras, 438, 3340

\bibitem[{{Felton}(1999)}]{1999Felton}
{Felton}, M. 1999, {Optical polarimetry studies of Seyfert galaxies. Doctoral
  thesis, Durham University.}

\bibitem[{{Gaskell}(2009)}]{2009Gaskell}
{Gaskell}, C.~M. 2009, \nar, 53, 140

\bibitem[{{Gaskell} {et~al.}(2012){Gaskell}, {Goosmann}, {Merkulova},
  {Shakhovskoy}, \& {Shoji}}]{2012Gaskell}
{Gaskell}, C.~M., {Goosmann}, R.~W., {Merkulova}, N.~I., {Shakhovskoy}, N.~M.,
  \& {Shoji}, M. 2012, \apj, 749, 148

\bibitem[{{Gaskell} \& {Harrington}(2018)}]{2018Gaskell}
{Gaskell}, C.~M. \& {Harrington}, P.~Z. 2018, \mnras, 478, 1660

\bibitem[{{Goodrich}(1989)}]{1989Goodrich}
{Goodrich}, R.~W. 1989, \apj, 340, 190

\bibitem[{{Gregory} {et~al.}(1991){Gregory}, {Tifft}, \& {Cocke}}]{1991Gregory}
{Gregory}, S.~A., {Tifft}, W.~G., \& {Cocke}, W.~J. 1991, \aj, 102, 1977

\bibitem[{{Hutsem{\'e}kers} {et~al.}(2019){Hutsem{\'e}kers}, {Ag{\'\i}s
  Gonz{\'a}lez}, {Marin}, {Sluse}, {Ramos Almeida}, \& {Acosta
  Pulido}}]{2019Hutsemekers}
{Hutsem{\'e}kers}, D., {Ag{\'\i}s Gonz{\'a}lez}, B., {Marin}, F., {et~al.}
  2019, \aap, 625, A54

\bibitem[{{Hutsem{\'e}kers} {et~al.}(2017){Hutsem{\'e}kers}, {Ag{\'\i}s
  Gonz{\'a}lez}, {Sluse}, {Ramos Almeida}, \& {Acosta
  Pulido}}]{2017Hutsemekers}
{Hutsem{\'e}kers}, D., {Ag{\'\i}s Gonz{\'a}lez}, B., {Sluse}, D., {Ramos
  Almeida}, C., \& {Acosta Pulido}, J.~A. 2017, \aap, 604, L3

\bibitem[{{Impey} {et~al.}(1991){Impey}, {Lawrence}, \& {Tapia}}]{1991Impey}
{Impey}, C.~D., {Lawrence}, C.~R., \& {Tapia}, S. 1991, \apj, 375, 46

\bibitem[{{Khachikian} {et~al.}(2011){Khachikian}, {Asatrian}, \&
  {Burenkov}}]{2011Khachikian}
{Khachikian}, E.~Y., {Asatrian}, N.~S., \& {Burenkov}, A.~N. 2011,
  Astrophysics, 54, 26

\bibitem[{{Khachikian} \& {Weedman}(1971)}]{1971Khachikian}
{Khachikian}, E.~Y. \& {Weedman}, D.~W. 1971, \apjl, 164, L109

\bibitem[{{Kollatschny} {et~al.}(2000){Kollatschny}, {Bischoff}, \&
  {Dietrich}}]{2000Kollatschny}
{Kollatschny}, W., {Bischoff}, K., \& {Dietrich}, M. 2000, \aap, 361, 901

\bibitem[{{Kollatschny} \& {Fricke}(1985)}]{1985Kollatschny}
{Kollatschny}, W. \& {Fricke}, K.~J. 1985, \aap, 146, L11

\bibitem[{{Lawrence} {et~al.}(2016){Lawrence}, {Bruce}, {MacLeod}, {Gezari},
  {Elvis}, {Ward}, {Smartt}, {Smith}, {Wright}, \& {Fraser}}]{2016Lawrence}
{Lawrence}, A., {Bruce}, A.~G., {MacLeod}, C., {et~al.} 2016, \mnras, 463, 296

\bibitem[{{Lub} \& {de Ruiter}(1992)}]{1992Lub}
{Lub}, J. \& {de Ruiter}, H.~R. 1992, \aap, 256, 33

\bibitem[{{Malkov} {et~al.}(1997){Malkov}, {Pronik}, \& {Sergeev}}]{1997Malkov}
{Malkov}, Y.~F., {Pronik}, V.~I., \& {Sergeev}, S.~G. 1997, \aap, 324, 904

\bibitem[{{Marin}(2017)}]{2017Marin}
{Marin}, F. 2017, \aap, 607, A40

\bibitem[{{Marin} {et~al.}(2016){Marin}, {Goosmann}, \& {Petrucci}}]{2016Marin}
{Marin}, F., {Goosmann}, R.~W., \& {Petrucci}, P.~O. 2016, \aap, 591, A23

\bibitem[{{Martel}(1998)}]{1998Martel}
{Martel}, A.~R. 1998, \apj, 508, 657

\bibitem[{{Martin} {et~al.}(1983){Martin}, {Thompson}, {Maza}, \&
  {Angel}}]{1983Martin}
{Martin}, P.~G., {Thompson}, I.~B., {Maza}, J., \& {Angel}, J.~R.~P. 1983,
  \apj, 266, 470

\bibitem[{{McElroy} {et~al.}(2016){McElroy}, {Husemann}, {Croom}, {Davis},
  {Bennert}, {Busch}, {Combes}, {Eckart}, {Perez-Torres}, {Powell},
  {Scharw{\"a}chter}, {Tremblay}, \& {Urrutia}}]{2016McElroy}
{McElroy}, R.~E., {Husemann}, B., {Croom}, S.~M., {et~al.} 2016, \aap, 593, L8

\bibitem[{{Moran} {et~al.}(1996){Moran}, {Halpern}, \& {Helfand}}]{1996Moran}
{Moran}, E.~C., {Halpern}, J.~P., \& {Helfand}, D.~J. 1996, \apjs, 106, 341

\bibitem[{{Noda} \& {Done}(2018)}]{2018Noda}
{Noda}, H. \& {Done}, C. 2018, \mnras, 480, 3898

\bibitem[{{Oknyansky} {et~al.}(2017){Oknyansky}, {Gaskell}, {Huseynov},
  {Lipunov}, {Shatsky}, {Tsygankov}, {Gorbovskoy}, {Mikailov}, {Tatarnikov},
  {Buckley}, {Metlov}, {Nadzhip}, {Kuznetsov}, {Balanutza}, {Burlak},
  {Galazutdinov}, {Artamonov}, {Salmanov}, {Malanchev}, \&
  {Oknyansky}}]{2017Oknyansky}
{Oknyansky}, V.~L., {Gaskell}, C.~M., {Huseynov}, N.~A., {et~al.} 2017, \mnras,
  467, 1496

\bibitem[{{Oknyansky} {et~al.}(2019){Oknyansky}, {Winkler}, {Tsygankov},
  {Lipunov}, {Gorbovskoy}, {van Wyk}, {Buckley}, \& {Tyurina}}]{2019Oknyansky}
{Oknyansky}, V.~L., {Winkler}, H., {Tsygankov}, S.~S., {et~al.} 2019, \mnras,
  483, 558

\bibitem[{{Pastoriza} \& {Gerola}(1970)}]{1970Pastoriza}
{Pastoriza}, M. \& {Gerola}, H. 1970, Astrophysical Letters, 6, 155

\bibitem[{{Penston} \& {Perez}(1984)}]{1984Penston}
{Penston}, M.~V. \& {Perez}, E. 1984, \mnras, 211, 33P

\bibitem[{{Raimundo} {et~al.}(2019){Raimundo}, {Vestergaard}, {Koay},
  {Lawther}, {Casasola}, \& {Peterson}}]{2019Raimundo}
{Raimundo}, S.~I., {Vestergaard}, M., {Koay}, J.~Y., {et~al.} 2019, \mnras,
  486, 123

\bibitem[{{Rees}(1988)}]{1988Rees}
{Rees}, M.~J. 1988, \nat, 333, 523

\bibitem[{{Sergeev} {et~al.}(2017){Sergeev}, {Nazarov}, \&
  {Borman}}]{2017Sergeev}
{Sergeev}, S.~G., {Nazarov}, S.~V., \& {Borman}, G.~A. 2017, \mnras, 465, 1898

\bibitem[{{Shapovalova} {et~al.}(2019){Shapovalova}, {Popovi{\'c}}, {},
  {Afanasiev}, {Ili{\'c}}, {}, {Kova{\v c}evi{\'c}}, {}, {Burenkov},
  {Chavushyan}, {Mar{\v c}eta-Mandi{\'c}}, {}, {Spiridonova}, {Valdes},
  {Bochkarev}, {Pati{\~n}o-{\'A}lvarez}, {Carrasco}, \&
  {Zhdanova}}]{2019Shapovalova}
{Shapovalova}, A.~I., {Popovi{\'c}}, {}, L.~{\v C}., {et~al.} 2019, \mnras,
  485, 4790

\bibitem[{{Shapovalova} {et~al.}(2008){Shapovalova}, {Popovi{\'c}}, {Collin},
  {Burenkov}, {Chavushyan}, {Bochkarev}, {Ben{\'\i}tez}, {Dultzin},
  {Kova{\v{c}}evi{\'c}}, \& {Borisov}}]{2008Shapovalova}
{Shapovalova}, A.~I., {Popovi{\'c}}, L.~{\v{C}}., {Collin}, S., {et~al.} 2008,
  \aap, 486, 99

\bibitem[{{Shappee} {et~al.}(2014){Shappee}, {Prieto}, {Grupe}, {Kochanek},
  {Stanek}, {De Rosa}, {Mathur}, {Zu}, {Peterson}, {Pogge}, {Komossa}, {Im},
  {Jencson}, {Holoien}, {Basu}, {Beacom}, {Szczygie{\l}}, {Brimacombe},
  {Adams}, {Campillay}, {Choi}, {Contreras}, {Dietrich}, {Dubberley},
  {Elphick}, {Foale}, {Giustini}, {Gonzalez}, {Hawkins}, {Howell}, {Hsiao},
  {Koss}, {Leighly}, {Morrell}, {Mudd}, {Mullins}, {Nugent}, {Parrent},
  {Phillips}, {Pojmanski}, {Rosing}, {Ross}, {Sand}, {Terndrup}, {Valenti},
  {Walker}, \& {Yoon}}]{2014Shappee}
{Shappee}, B.~J., {Prieto}, J.~L., {Grupe}, D., {et~al.} 2014, \apj, 788, 48

\bibitem[{{Smith} {et~al.}(2004){Smith}, {Robinson}, {Alexander}, {Young},
  {Axon}, \& {Corbett}}]{2004Smith}
{Smith}, J.~E., {Robinson}, A., {Alexander}, D.~M., {et~al.} 2004, \mnras, 350,
  140

\bibitem[{{Smith} {et~al.}(2002){Smith}, {Schmidt}, {Hines}, {Cutri}, \&
  {Nelson}}]{2002Smith}
{Smith}, P.~S., {Schmidt}, G.~D., {Hines}, D.~C., {Cutri}, R.~M., \& {Nelson},
  B.~O. 2002, \apj, 569, 23

\bibitem[{{Tohline} \& {Osterbrock}(1976)}]{1976Tohline}
{Tohline}, J.~E. \& {Osterbrock}, D.~E. 1976, \apj, 210, L117

\bibitem[{{Tran} {et~al.}(1992){Tran}, {Osterbrock}, \& {Martel}}]{1992Tran}
{Tran}, H.~D., {Osterbrock}, D.~E., \& {Martel}, A. 1992, \aj, 104, 2072

\bibitem[{{Veilleux} \& {Zheng}(1991)}]{1991Veilleux}
{Veilleux}, S. \& {Zheng}, W. 1991, \apj, 377, 89

\bibitem[{{Ward} {et~al.}(1980){Ward}, {Penston}, {Blades}, \&
  {Turtle}}]{1980Ward}
{Ward}, M., {Penston}, M.~V., {Blades}, J.~C., \& {Turtle}, A.~J. 1980, \mnras,
  193, 563

\bibitem[{{Wills} {et~al.}(2011){Wills}, {Wills}, \& {Breger}}]{2011Wills}
{Wills}, B.~J., {Wills}, D., \& {Breger}, M. 2011, \apjs, 194, 19

\end{thebibliography}

\end{document}